# Perpendicular magnetic anisotropy and spin glass-like behavior in molecular beam epitaxy grown chromium telluride thin films

Anupam Roy [*], Samaresh Guchhait, Rik Dey, Tanmoy Pramanik, Cheng-Chih Hsieh, Amritesh Rai and Sanjay K Banerjee

Microelectronics Research Center, The University of Texas at Austin, Austin, Texas 78758, USA

*Address correspondence to anupam@austin.utexas.edu

**ABSTRACT**: Reflection high energy electron diffraction (RHEED), scanning tunneling microscopy (STM), vibrating sample magnetometry and other physical property measurements are used to investigate the structure, morphology, magnetic and magneto-transport properties of (001)-oriented $Cr_2Te_3$ thin films grown on $Al_2O_3$(0001) and Si(111)-(7×7) surfaces by molecular beam epitaxy (MBE). Streaky RHEED patterns indicate flat smooth film growth on both substrates. STM studies show the hexagonal arrangements of surface atoms. Determination of the lattice parameter from atomically resolved STM image is consistent with the bulk crystal structures. Magnetic measurements show the film is ferromagnetic having the Curie temperature of about 180 K, and a spin glass-like behavior was observed below 35 K. Magneto-transport measurements show the metallic nature of the film with a perpendicular magnetic anisotropy along the *c*-axis.

Transition-metal chalcogenides have attracted much interest in the past due to their large variety of novel physical, electrical and magnetic properties. On the one hand, these can be metallic, half-metallic or semiconducting depending on the anion, and on the ratio between the transition metal and chalcogen atoms. On the other hand, these chalcogenides may have ferromagnetic, antiferromagnetic or non-collinear spin structure for different crystal structures and compositions. Chromium-tellurium system has a large family of compounds. There are various stable stoichiometries [*e.g.*, $Cr_{1-x}Te$, $Cr_2Te_3$, $Cr_3Te_4$, $Cr_5Te_8$, *etc.*] depending on the Cr-vacancies that occur in every second metal layer. Thus Cr-deficient and Cr-full layers stack alternatively along the *c*-axis.[1] All of these chromium chalcogenides have NiAs-type crystal structures and the stable phases are ferromagnetic, with a wide range of Curie temperatures, $T_c$, between 180 to 340 K. $T_c$ depends very sensitively on the composition of the compound. Wontcheu *et al.*[1] have shown the effect of anion substitution on the structural and magnetic properties of chromium chalcogenides. The physical properties change drastically with the change in composition.

Because of its unusual magnetic properties, $Cr_2Te_3$ is one of the interesting compounds in this family. Recently, Akiyama *et al.* have used thin $Cr_2Te_3$ ferromagnetic metallic film in a field-effect capacitor (FEC) structure.[2] Saito *et al.* have studied tunneling magnetoresistance (TMR) in the magnetic tunneling junctions (MTJs) with $Cr_{1-\delta}Te$ being one of the electrodes.[3] Several groups have studied the electronic and magnetic structures of $Cr_2Te_3$ bulk samples theoretically[4, 5] as well as experimentally.[1, 6-11] There are only a few reports of the epitaxial growth of $Cr_{1-\delta}Te$ thin films on GaAs(001) substrates.[2, 12-15] But, a detailed study of the structure, magnetic and transport properties of $Cr_2Te_3$ thin films is lacking. We have studied the growth of $Cr_2Te_3$ thin films directly on Si(111)-(7×7) and $Al_2O_3$(0001) surfaces using molecular beam epitaxy (MBE). We show the atomically resolved scanning tunneling microscopy (STM) micrographs of $Cr_2Te_3$ grown on Si(111)-(7×7) surfaces. Magnetic studies confirm the film to be ferromagnetic with a spin glass-like behavior at low temperature. We also have observed the anisotropic magnetoresistance (AMR) on the grown film. AMR is a well-known phenomenon observed in ferromagnetic materials with metallic conductance where the resistance changes with the angle between the current flow direction and the magnetization direction.[16] Interestingly, the grown films possess perpendicular magnetic anisotropy (PMA). Ferromagnetic PMA thin films have attracted interest due to their interesting fundamental properties and technological applications in magnetic recording,[17, 18] MTJs[19] and spin-transfer torque (STT) devices.[20, 21] Details of growth, structural, magnetic and transport properties are presented in the paper.

**RESULTS AND DISCUSSION**
**Growth and structural properties**

Previous report of growth of $Cr_{1-\delta}Te$ using MBE was on GaAs(001) substrates with buffer layers of ZnTe and CdTe.[2, 12-15] Here, we present the MBE growth of $Cr_2Te_3$ thin film of different thicknesses directly on UHV-cleaned $Al_2O_3$(0001) and Si(111)-(7×7) substrates without any additional buffer layer. *In situ* reflection high-energy electron diffraction (RHEED) study, monitored during the MBE growth of $Cr_2Te_3$ films grown on $Al_2O_3$(0001) and Si(111)-(7×7) substrates, is shown in Figure 1. Figures 1a and 1b show the RHEED patterns from a clean $Al_2O_3$(0001) substrate surface along $[1\ 0\ -1\ 0]_{Al2O3}$ and $[1\ 1\ -2\ 0]_{Al2O3}$ electron beam incidence, respectively. Corresponding RHEED patterns from the same surface following growth of $Cr_2Te_3$ thin films are shown in Figures 1c and 1d. Similar smooth film growth has been also achieved on Si(111) substrates. RHEED patterns of a reconstructed Si(111)-(7×7) surface are shown in Figure 1e for the electron beam along $[1\ 1\ -2]_{Si}$ direction and in Figure 1f for $[1\ -1\ 0]_{Si}$ incidence. Corresponding RHEED patterns from $Cr_2Te_3$ film show sharp streaky features in Figure 1g and 1h. This is evident that, on both the substrates, $Cr_2Te_3$ grows with a high crystalline quality giving atomically flat surface morphologies. Several samples with different thicknesses prepared on both the substrates show similar RHEED patterns and the RHEED patterns were maintained throughout the entire growth process. The growth is along (001) direction (also evident from XRD), which is very consistent for the growth of a hexagonal thin film on hcp(0001) or fcc(111) substrates. Similar (001)-oriented hexagonal thin film growth on fcc(111) structure has also been reported for $Bi_2Te_3$(0001) on Si(111) surfaces.[22]

Figure 2a shows the X-ray diffraction (XRD) pattern from a 12 nm thin film of $Cr_2Te_3$ grown on $Al_2O_3$(0001) substrate. The diffraction pattern matches very well with the NiAs-type hexagonal structure with the P-31*c* (163) space group. Peaks from (004) and (008) planes of $Cr_2Te_3$ film are indexed in the Figure 2a. It clearly shows that the grown film is following the underlying crystal symmetry of the substrate and growing along (001) direction. XRD pattern also rules out any significant presence of any impurities or other known phases of chromium telluride. *In situ* X-ray photoelectron spectroscopy (XPS) measurement from 12 nm thick $Cr_2Te_3$ film grown on $Al_2O_3$(0001) substrate shows the peaks correspond to Cr and Te. Figure 2b shows Cr-2*p* and Te-3*d* peaks. Since the binding energies of these two peaks are very close to each other, it is difficult to separate them out. The positions of the peaks are consistent with the reported results.[23]

A detailed microscopy study of $Cr_2Te_3$ thin film has been lacking, as previous studies do not present any high-resolution microscopy study. Sreenivasan *et al.*[14] have studied the roughness of the grown film by *ex situ* atomic force microscopy in a length scale which cannot reveal the surface structure at the atomic level. In Figure 3, we present an extensive *in situ* STM study of the surface of $Cr_2Te_3$ thin film grown on Si(111)-(7×7) surfaces. As observed from Figure 3a, the structures are characteristically triangular shaped, reflecting the hexagonal crystal structure along the (001) direction. This is not surprising as both hcp(0001) and fcc(111) surfaces have hexagonal Bravais lattice and differ only in registry of the third layer.[24] Because of the surface symmetry, formation of equilateral triangles on Si(111) substrate is quite natural. Triangle-shaped structures and terraces have also been observed for the growth of hexagonal $Bi_2Se_3$(0001) on Si(111) surfaces.[25, 26] Formation of different shapes of structures, including triangular and hexagonal on a surface of threefold symmetry have been shown by kinetic Monte Carlo simulation studies.[27, 28] The hexagonal structures on fcc(111) surfaces can have two types of edges. Final shape of the grown structure is determined by the competition of advancement rate of these two types of edges.[27, 28] Careful investigation reveals that, in our case, structures are truncated triangular (or, more precisely, triangular hexagon) in shape [Figure 3b and 3c], as should be the case when one type of edge advances faster than the other edge. Similar triangular hexagon structures have also been observed for the growth of $CoSi_2$ on Si(111) surfaces.[29] STM measurements carried out on several $Cr_2Te_3$ thin films of different thicknesses (4 nm to 20 nm) show similar triangular hexagons and hexagonal surface lattices indicating the growth following strictly the underlying crystal symmetry.

Figure 3a also shows the spirals and depressions in the grown film. One such spiral on a triangular hexagon structure is shown in Figure 3c. Spiral growth mode is observed for crystals with atomically flat surfaces. Such crystals grow by adatom incorporation at monatomic steps. Spiral growth mode is very common for GaN growth on SiC(0001) and $Al_2O_3$(0001) substrates where the origin of spiral growth has been determined to be high density threading dislocations with screw component.[30] Pioneering theoretical work by Burton, Cabrera and Frank (BCF) shows that a screw dislocation emerging from a crystal provides a continuous step source on the surface leading to the formation of growth spirals.[31] During growth this step winds around the dislocation center and thus forms a spiral. For $Cr_2Te_3$ thin film grown on Si(111) substrates, the spiral density is observed to be much lower compared to GaN growth. All the spirals observed are single arm spirals only. Any cooperative spirals with more than one arm are absent. Cui *et al.*[30] discussed the formation and annihilation of several types of spirals and different interactions between them. A spiral will be stable if the curvature of the spiral is more than a critical value. These spirals can be rotating clockwise and counter clockwise. Figure 3a shows that both types of rotations are present in $Cr_2Te_3$ spirals. Two such spirals present on two different triangular structures are shown (marked as "1" and "2") to have opposite rotations. Structure marked as "3" contains two spirals of opposite sign on the same triangular structure. Figure 3c also shows one of such instances where two spirals are on the same truncated hexagon.

Figure 3d shows the atomically resolved structure of $Cr_2Te_3$ surface. Hexagonal lattice arrangement on the surface is observed at room temperature (RT). In Figure 3e, the fast Fourier transformed (FFT) pattern from the image in Figure 3d, shows the periodic spots corresponding to the $Cr_2Te_3$ hexagonal surface symmetry. Inset shows the FFT-filtered STM image clearly indicating the hexagonal arrangement of surface atoms. One such hexagonal unit cell is marked (inset). Line profile drawn along one side of the marked hexagonal unit (arrow marked) is shown in Figure 3f. Lattice parameter *a*, from the line profile (marked as two vertical lines) is determined to be 6.68 Å which is very close to the corresponding bulk $Cr_2Te_3$ lattice parameter (6.81 Å) at RT as measured from the neutron diffraction study.[8] Lattice parameter obtained from the STM measurement also agrees very well with the measurement from transmission electron diffraction study as shown in ref.[12]

**Transport and Magnetic properties**

Figure 4a and 4b show the electrical resistivity measured from 300 K down to 2 K. The measurements were conducted using standard van der Pauw geometry on $Cr_2Te_3$ thin film grown on $Al_2O_3$(0001) and Si(111) substrates of thicknesses 4 nm and 20 nm, respectively. Both the curves show metallic behavior for the entire temperature range. The slopes of the electrical resistivity curves change abruptly at ~180 K which corresponds to the magnetic phase transition temperature. Magnetic measurement [Figure 4c] also shows that the Curie temperature ($T_c$) of the film is ~180 K. Below the Curie temperature, the electrical resistivity decreases more rapidly with temperature than above the Curie temperature. This may be because of reduction in electron-magnon scattering due to ferromagnetic spin alignment, which is explained next. Electrical resistivity curve for $Cr_{1.96}Te_3$ bulk samples also shows a transition point around similar temperature range.[10] Similar trend in resistivity is also observed for ferromagnetic transition metals, where the resistivity of a ferromagnetic metal can be described as $\varrho(T) = \varrho_0 + \varrho_L(T) + \varrho_M(T)$. Here $\varrho_0$ is the residual resistivity at absolute zero arising due to the scattering of electrons from lattice defects and impurities and $\varrho_L(T)$ comes from the scattering of conduction electrons by the lattice vibrations (phonons), which increases with temperature. The additional $\varrho_M(T)$ term arises only for the ferromagnetic materials as the electrons get scattered by magnons below $T_c$.[32] The magnetic scattering arises from the *s-d* exchange interaction between the conduction electrons and the more localized 3*d* magnetic electrons.[33] Above $T_c$, the spins are disordered and the resistance due to scattering related to magnetic order approaches a temperature-independent saturation value. However, below $T_c$, spontaneous magnetization appears, which aligns spins along the magnetization direction. Ferromagnetic alignment of spins reduces the electron-magnon scattering. When temperature is lowered, more spins align ferromagnetically [Figure 4c], and this leads to further reduction of electron-magnon scattering with decreasing temperature. Careful observation also reveals that the percentage change of resistance below $T_c$ is lower than that observed in typical transition metals.[34] That is probably due to the presence of magnetic domains and their freeze out in random direction with decreasing temperature. Electrons get scattered at domain boundaries and hence the reduction of resistance is smaller than a bulk, uniformly magnetized sample.

Magnetic measurements from 20 nm $Cr_2Te_3$ film grown on Si(111) surfaces have been shown in Figure 4c and 4d. Figure 4c shows the variation of the field-cooled (FC) and zero-field-cooled (ZFC) magnetization with temperature in a magnetic field of 500 Oe along the surface plane of $Cr_2Te_3$ thin film. Field-cooled magnetization ($M_{FC}$) shows a paramagnetic-to-ferromagnetic transition at $T_c = 180$ K and increases continuously with decreasing temperature below $T_c$. A strong continuous increase of $M_{FC}$ below $T_c$ indicates the ferromagnetic nature of the grown film. Reported $T_c$ for $Cr_2Te_3$ bulk samples (170-180 K) matches very well with our result.[4, 10, 11] Magnetic hysteresis curve at 2 K, shown in Figure 4d, also indicates the ferromagnetic nature of the film. The magnetic moment is determined to be ~$2.8\mu_B$ per Cr atom ($\mu_B$ is the Bohr magnetron). Itinerant nature of Cr 3*d* electrons appears to be the reason behind this discrepancy from the expected saturation magnetic moment of $3\mu_B$ per Cr atom calculated from an ionic model and also observed by others in bulk and thin film samples.[5, 9, 10, 12] The magnetization does not saturate completely even with a strong magnetic field [as observed in Figure 4d]. This non-saturation of magnetic moment has been observed for other transition metal compounds where frustration is present and also been reported before.[10, 12] Existence of a spin glass phase in our sample (discussed below) also shows presence of frustration in $Cr_2Te_3$ sample.

Another interesting feature of Figure 4c is that the zero-field-cooled magnetization ($M_{ZFC}$) first increases and then become almost constant with increasing temperature. The temperature where this abrupt change of slope takes place is defined as the freezing temperature $T_f$. On the contrary, $M_{FC}$ always decreases with increasing temperature below $T_c$. Furthermore, with further increase in temperature, $M_{ZFC}$ first decreases slightly and then merges with $M_{FC}$ at a temperature $T_b$ just below $T_c$. Temperature $T_b$ is also known as the blocking temperature. Above this temperature $T_b$, $M_{ZFC}$ and $M_{FC}$ superpose each other. The freezing temperature ($T_f$), an indication of a spin glass-like phase, is determined to be ~ 35 K. Also note that in Figures 4a and 4b, a noticeable change in slope of the resistivity curves is observed near 35 K. However, a large difference between $M_{FC}$ and $M_{ZFC}$ values between $T_f$ and $T_b$ indicates the presence of short range ferromagnetic ordering (*i.e.*, domains) within our sample. Freezing out of domains in random direction in absence of any magnetic field can explain the observed difference between $M_{FC}$ and $M_{ZFC}$ magnetization curves,[35] as well as transport properties of our samples below $T_c$.

Spin glass-like behavior has also been reported for $Cr_7(Se_{1-x}Te_x)_8$ compounds before, resulting from the spin frustration due to the competition between the ferromagnetic and antiferromagnetic interactions.[1, 36] Hui *et al*.[12] have suggested that antiferromagnetically coupled $Cr^{3+}$ spins in the vacancy layer of NiAs-type structure[4, 12] may have a role in the observed intrinsic magnetic exchange bias in structurally single crystal $Cr_2Te_3$ film. As stated before, between $T_c$ and $T_f$, $M_{ZFC}$ is lower than $M_{FC}$ due to the freeze out of the magnetic domains in random direction with decreasing temperature. However, we believe a spin glass-like phase appears below $T_f$ that causes more randomization of magnetic moments within each domain and hence $M_{ZFC}$ drops more sharply with decreasing temperature.

Since with decreasing temperature here we have paramagnetic-to-ferromagnetic to spin glass transitions, we believe this might be a re-entrant type spin glass phase. A re-entrant spin glass phase arises when both (stronger) ferromagnetic and (weaker) anti-ferromagnetic interactions are present within the same system. In such a situation, one will first see a ferromagnetic transition. However, for a "strong enough" anti-ferromagnetic interaction, the system becomes frustrated at lower temperatures and hence a spin glass transition may happen at even lower temperature.[37] Hashimoto *et al*.[10] have claimed that both stronger ferromagnetic and weaker anti-ferromagnetic Cr-Cr interactions are present within $Cr_2Te_3$ and hence they have predicted the existence of more "complex magnetic order below the Curie temperature". We believe this may explain the origin of re-entrant spin glass phase in our $Cr_2Te_3$ sample. Also note that non-saturation of magnetic moment at very large magnetic field at 2 K [Figure 4d] is a clear indication of frustrated interactions in this system. Similar coexistence of ferromagnetism and spin glass behavior has been observed in many systems, including amorphous Ge:Mn[38] and a cluster-glass perovskite compound where the cusp in $M_{ZFC}$ is governed by a local anisotropy field acting on the spins inside each domain.[39] This cluster-glass phase is nothing but a modification to the spin-glass system formed when the magnetic spin density is increased due to short range ferromagnetic ordering and thus resulting in formation of magnetic clusters.[1, 40, 41] However, more studies are necessary to understand the true nature of the spin glass phase in $Cr_2Te_3$ and distinguish between different glassy systems [*e.g.*, cluster-glass, reentrant spin-glass, canted spin system *etc*.] that can show many common macroscopic features.[40] This coexistence of ferromagnetism and re-entrant glassy behavior is an interesting new observation in $Cr_2Te_3$ thin film.

Figure 5a shows the variation of magnetoresistance (MR) at 2 K for the $Cr_2Te_3$ film grown on $Al_2O_3$(0001) surfaces with the magnetic field oriented parallel ($\theta = 90°$) and perpendicular ($\theta = 0°$) to the surface plane. From parallel and perpendicular field MR, we conclude that $Cr_2Te_3$ film possesses a perpendicular magnetic anisotropy (PMA), which is explained next. At zero field value, magnetic easy axis is normal to the surface and so is the magnetization direction. The resistance at zero field is minimum because the magnetization direction is normal to the current flow direction. With a magnetic field applied parallel to the surface, the effective magnetization direction starts rotating towards the in-plane magnetic field until it saturates. At this saturation field (parallel to the surface), the magnetization reaches its saturation magnetization value and points towards the direction of current flow and lies along the surface. Hence at this saturation field (~ 4 T at 2 K), the resistance also becomes maximum. This field dependence of MR corresponds to the anisotropic MR (AMR), which results from the anisotropy of spin-orbit interaction in ferromagnetic materials.[16] For higher field values, the obtained MR is linear and shows negative slope, which is due to the suppression of weak localization (WL) and/or electron-electron interaction (EEI).[42-44] Hashimoto et al.[10] have studied the magnetization of $Cr_2Te_3$ bulk crystal and observed that the magnetic easy axis is pointing along the c-axis. As we have observed from RHEED, XRD and the STM studies, $Cr_2Te_3$ grows epitaxially along the c-axis on $Al_2O_3$(0001) and Si(111) substrates. PMA observed from transport studies [Figure 5a] thus indicates a magnetic crystalline anisotropy present in the 4 nm thick $Cr_2Te_3$ epitaxial film.

When the magnetic field applied perpendicular to the surface, the MR consists of two parts - one is almost linear, reversible negative MR at high field that arises due to the suppression of WL and/or EEI. At smaller field range, MR shows a hysteresis with back-and-forth field sweep and has a sharp maximum. The two sharp maxima in the resistance correspond to the coercive field value (~1 T) where the domain wall density reaches its maximum (*i.e.*, the domains are all misaligned), and as a result, the resistance is also maximum due to increased scattering at domain boundaries. As the field increases the domains start to align in the direction of the field and the resistance starts to decrease. However, this negative MR does not saturate even at high field (9 T) and low temperature (2 K), where most spins are ferromagnetically coupled with each other, due to suppression of WL and/or EEI. It is interesting to note that the resistance values at zero fields are not the same for parallel and perpendicular field sweeps. As the film possesses PMA, in perpendicular field sweep more domains are aligned with each other at zero field; while in parallel field sweep most domains are misaligned due to a previously applied parallel field. This causes the resistance value at zero field to be smaller for the perpendicular field sweep than that for the parallel field sweep.

Presence of PMA is also evident from the variation of MR with the magnetic field applied at different angles with respect to the surface normal (as shown in Figure 5b). Except for the parallel field sweep, the resistance value at zero field for the field sweeping in different orientations are almost the same. As long as there is some component of magnetic field in the perpendicular direction, the magnetization aligns along the easy axis of this PMA film and shows almost same value of resistance at zero field. However, for the parallel field orientation, the magnetization does not align along the easy axis due to a previously applied parallel field and the resistance is higher at the zero field due to the maximum randomness in domain alignment. Also the magnetic field at which the peak appears in the MR increases with the change in field orientation from $\theta = 0°$ (perpendicular field) to $\theta = 90°$ (parallel field). As the field direction changes from the perpendicular to parallel, higher magnetic field is necessary to attain the maximum randomness. So it is clear that for a field oriented between 0° and 90°, the MR is due to the competition between the components of magnetic field perpendicular and parallel to the surface. Figure 5c shows the parallel field MR at different temperatures. The saturation field value, where the slope of MR changes, decreases with increasing temperature, as expected for a ferromagnetic film. Figure 5d shows the perpendicular field MR at different temperatures. The coercive field value, where the sharp maximum in the MR occurs, decreases with increasing temperature, which is also a characteristic of a ferromagnetic film.[45] As temperature increases, domains can more easily align along the applied field and as a result coercive field value is also smaller.

In several applications, materials with PMA are used for superior performances.[17-21] PMA is observed in very thin transition metal layers, such as 0.8 nm or less thick Co layer. Multilayers of transition metals with other non-magnetic metals, such as Co/Pd multilayers, are used in MTJ type devices. These type of PMA multilayers are used in MTJ because of higher thermal stability, more magnetic uniformity and larger magnetic anisotropy energy compared to in-plane anisotropy materials. Moreover, for STT memory applications, PMA materials are useful due to lower switching current requirement.[46] However the presence of multi-interfaces increases the overall resistance of the film. A single layer PMA film of $Cr_2Te_3$ can reduce the number of interfaces and thus reducing the overall loss of spin polarization due to inelastic scattering at the interfaces. Considering the saturation magnetization, $M_S$ of $620 \times 10^3$ A/m (from Figure 4d) and the anisotropy field $H_A$ of 4.1 T (from Figure 5a), the PMA energy density, $E_p = -M_S H_A/2$, is estimated approximately to be $1.27 \times 10^6$ J/m$^3$. The estimated value is in reasonably good comparison with other PMA films.[47] As all the chromium telluride compounds are of the same NiAs-type crystal structure, one can expect to have a single-crystalline PMA film with a wide range of $T_c$ for various Cr compositions. We believe, this work opens up an opportunity to study the PMA property of chromium telluride films and provides an excellent possibility for various applications at a wide temperature range.

### SUMMARY AND CONCLUSIONS

In conclusion, we have carried out the MBE growth of $Cr_2Te_3$ thin films on $Al_2O_3(0001)$ and Si(111) substrates. Structural, magnetic and transport properties of the films have been characterized by several *in situ* and *ex situ* techniques. Sharp streaks in RHEED patterns imply smooth thin film growth on both the substrates. As observed from RHEED and XRD, the as-grown film is hexagonal and oriented along (001) direction (*c*-axis). We have shown the hexagonal atomic arrangement of $Cr_2Te_3$ film from high-resolution *in situ* STM measurements at room temperature. Magnetic measurement shows the film to be ferromagnetic and a spin glass-like phase appears below 35 K. This shows competing interactions within $Cr_2Te_3$. Magneto-transport studies reveal that the film possesses perpendicular magnetic anisotropy in a 4 nm $Cr_2Te_3$ film, which has not been observed before. Presence of PMA makes it a very useful material for possible spintronics applications.

### EXPERIMENTAL

$Cr_2Te_3$ films were grown in a custom-built MBE growth system (Omicron, Germany) under ultra-high vacuum (UHV) conditions (base pressure ~$1\times10^{-10}$ mbar). Detail of the system has been described elsewhere.[22] A RHEED setup is attached to the MBE system for *in situ* monitoring of surface reconstruction and growth. Substrates used in the experiment were insulating *c*-axis $Al_2O_3(0001)$ and P-doped n-type Si(111) wafers (oriented within ±0.5°) with a resistivity of 1-20 Ω-cm. Atomically clean, reconstructed Si(111)-(7×7) surfaces were prepared by the usual heating and flashing procedure[48] and *c*-Al$_2$O$_3$(0001) substrates were cleaned by the standard heating method in UHV. Clean substrate surfaces were examined by *in situ* RHEED. Chromium and tellurium fluxes generated by *e*-beam evaporator and effusion cell, respectively, were co-deposited onto the substrates at an elevated substrate temperature of about 340 °C. The chamber pressure during growth never exceeded 1×10$^{-9}$ mbar and the Te$_2$/Cr BEP (beam equivalent pressure) flux ratio was kept at about 15. Several samples with thicknesses varying from 4 nm to 20 nm were grown and typical growth rate of Cr$_2$Te$_3$ films was about 0.1 nm/min.

Post-growth investigations of the samples were carried out by *in situ* RHEED operated at 13 kV, STM at room temperature (RT) in the constant current mode, XPS with monochromatic Al-K*α* and *ex situ* XRD. Magnetic and transport measurements were carried out with 9 T Quantum Design physical property measurement system (PPMS) combined with vibrating sample magnetometry (VSM) capable of cooling samples down to ~ 2 K.

**Conflict of interest**: The authors declare no competing financial interest.

**Acknowledgment**: This work was supported by the NRI SWAN Center. We acknowledge helpful discussions with Professor A. H. MacDonald, Professor B. N. Dev, Dr. L. Colombo, Dr. Jiamin Xue and Dr. Sumalay Roy. We also appreciate technical support from Omicron and Quantum Design.

**FIGURE CAPTION**

**Figure 1.** RHEED images following $Cr_2Te_3$ growth on $Al_2O_3$(0001) and Si(111)-(7×7) surfaces. (a) and (b) RHEED patterns from a clean $Al_2O_3$(0001) surface with the incident electron beam along [1 0 -1 0] and [1 1 -2 0] orientations of $Al_2O_3$, respectively. (c) and (d) Corresponding RHEED patterns from the same surface following 4 nm of $Cr_2Te_3$ growth. (e) and (f) Typical (7×7) surface reconstruction from Si(111) substrate along [1 1 -2] and [1 -1 0] orientations of Si, respectively. (g) and (h) Corresponding RHEED patterns following 8 nm of $Cr_2Te_3$ growth.

**Figure 2.** (a) XRD pattern from 12 nm of epitaxial $Cr_2Te_3$ thin film. The pattern shows that the growth is along (001) direction. (b) Cr-2*p* and Te-3*d* core-level x-ray photoelectron spectra from 12 nm of $Cr_2Te_3$ thin film.

**Figure 3.** STM study of a 8 nm epitaxial $Cr_2Te_3$ thin film grown on Si(111)-(7×7) surfaces. (a) Several triangular features along with spirals having clockwise and anti-clockwise rotations. (Scan area: 500 × 500 $nm^2$, bias voltage: -1 V, tunneling current: 0.2 nA). (b) Truncated hexagon structure indicating the influence of substrate surface symmetry. (Scan area: 60 × 60 $nm^2$, bias voltage: -1 V, tunneling current: 0.2 nA). (c) Two spirals of opposite sign on a truncated hexagon. (Scan area: 70 × 70 $nm^2$, bias voltage: -1 V, tunneling current: 0.2 nA). (d) Atomically resolved STM image shows hexagonal units on the surface. (Scan area: 10 × 10 $nm^2$, bias

voltage: -1 V, tunneling current: 0.2 nA). (e) Fourier transformed pattern from the STM image in (d) shows diffraction spots corresponding to hexagonal unit cells. (Inset) FFT-filtered STM image shows hexagonal arrangement of surface atoms. One unit cell is marked. (f) Profile drawn across the line marked on the hexagonal unit in inset of (e).

**Figure 4.** Temperature dependence of electrical resistivity showing metallic behavior. (a) 4 nm and (b) 20 nm epitaxial $Cr_2Te_3$ film grown on $Al_2O_3$(0001) and Si(111)-(7 × 7) substrate surfaces, respectively. Magnetic studies from $Cr_2Te_3$ thin film grown on Si(111)-(7 × 7) surface. (c) Zero-field-cooled (ZFC) and field-cooled (FC) magnetization as a function of temperature with a 500 Oe magnetic field along the surface plane. (d) Hysteresis loops of $Cr_2Te_3$ thin film at 2 K with the magnetic field parallel to the surface plane showing non-saturating magnetization even with 5 T magnetic field. The inset shows an enlarged portion of the hysteresis loop around the origin.

**Figure 5.** Variation of magnetoresistance (MR) at 2 K with the magnetic field applied parallel ($\theta$ = 90°) and perpendicular ($\theta$ = 0°) to the surface. (b) Variation of MR at 2 K with respect to the magnetic field at different orientations. Variation of MR with the magnetic field applied (c) parallel and (d) perpendicular to the surface at different temperatures.

**Figure 1**

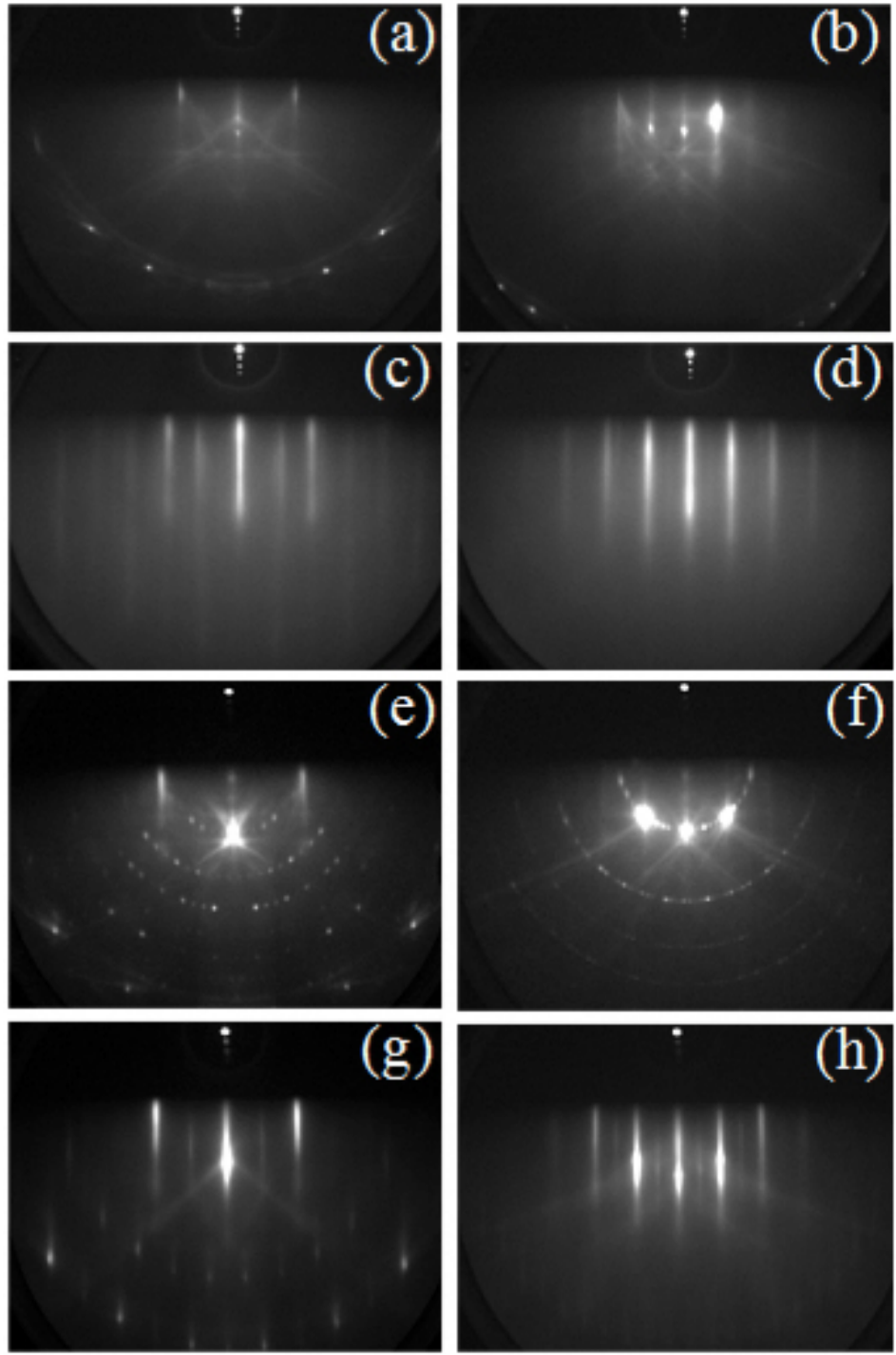

Figure 2

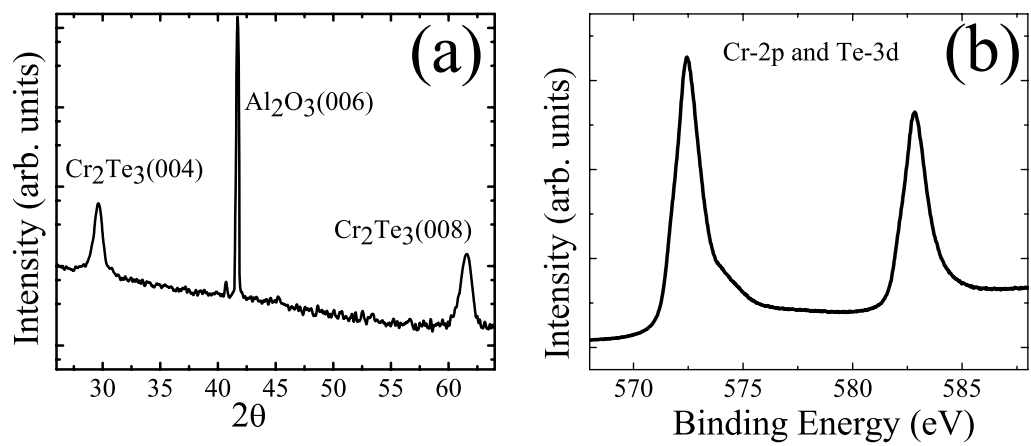

**Figure 3**

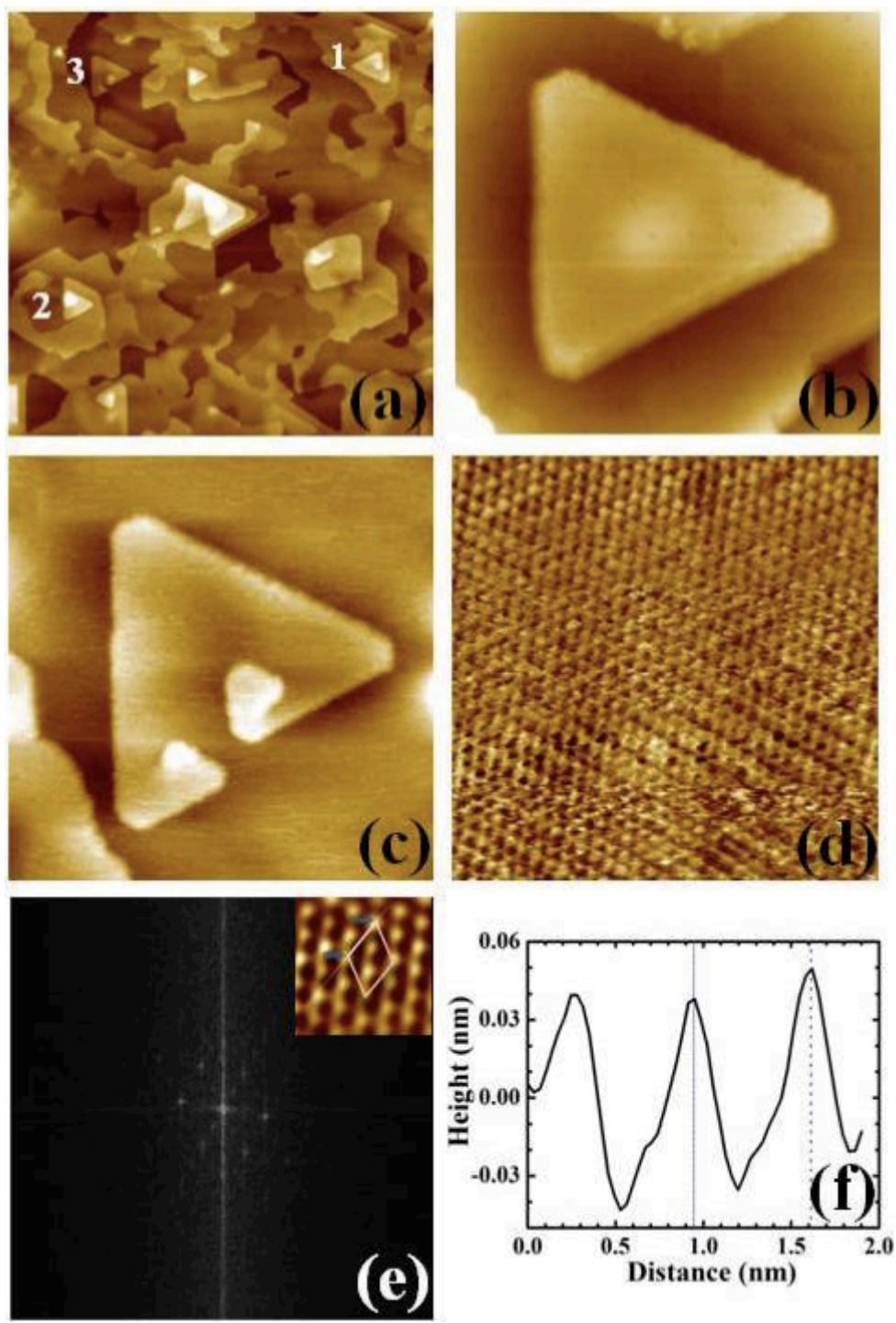

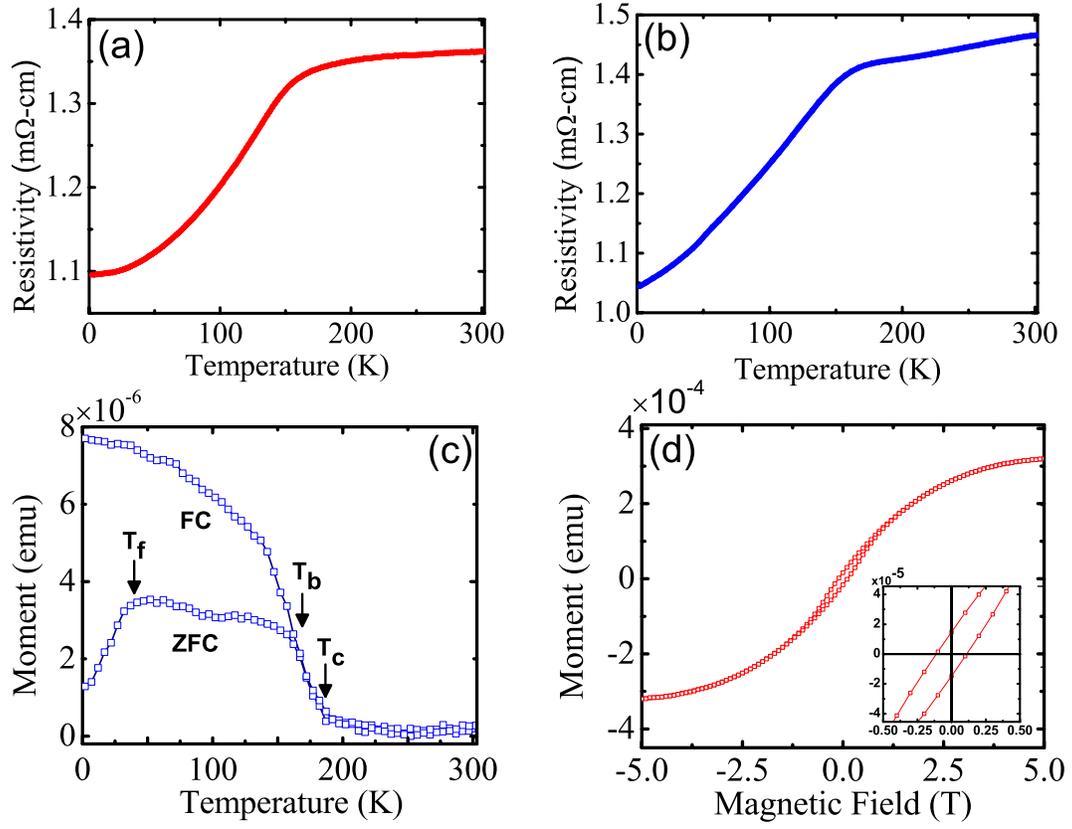

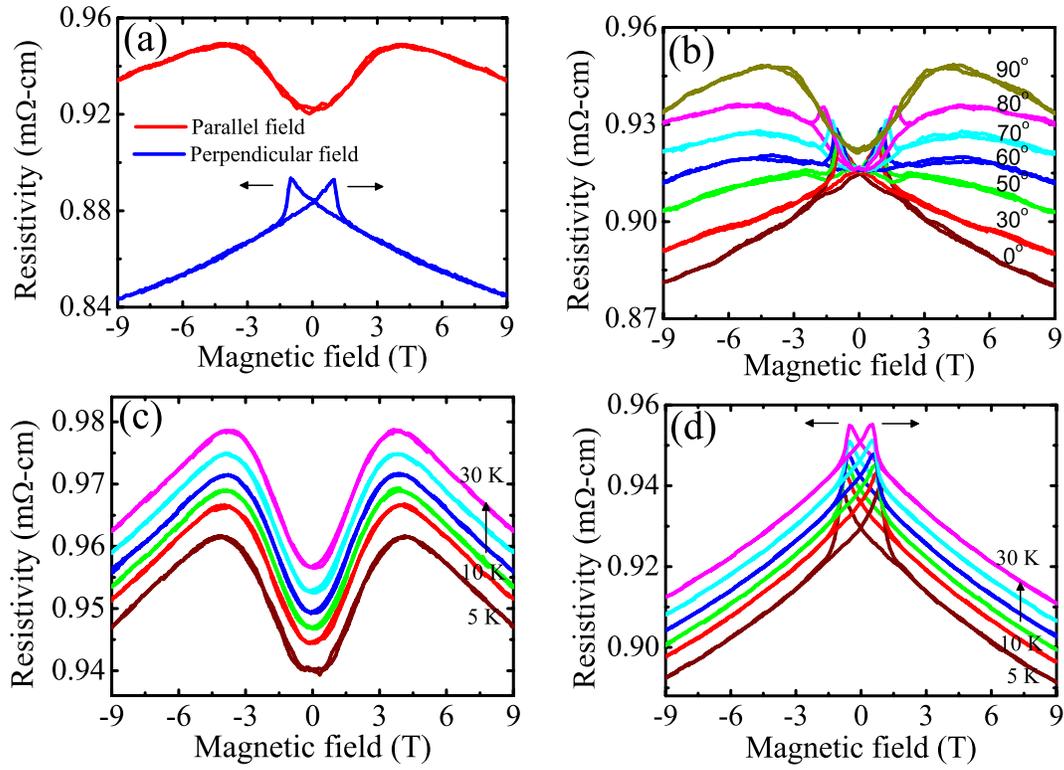